\documentclass[prb,showpacs,preprintnumbers,twocolumn,amsmath,amssymb,superscriptaddress]{revtex4}
\usepackage{graphicx}
\usepackage{color}
\usepackage{dcolumn}
\usepackage{bm}
\usepackage{psfrag}
\def\bq{{\bf q}}

\newcommand{\be}{\begin{equation}}
\newcommand{\ee}{\end{equation}}
\newcommand{\bea}{\begin{eqnarray}}
\newcommand{\eea}{\end{eqnarray}}

\def\a{\alpha}
\def\b{\beta}

\def\d{\delta}
\def\g{\gamma}

\def\o{\omega}

\def\s{\sigma}

\def\D{\Delta}

\def\L{\Lambda}

\def\ra{\rightarrow}

\def\bq{{\bf q}}

\def\nn{\nonumber}
\def\lb{\label}

\def\pref#1{(\ref{#1})}

\newcount\bozza \bozza=0
\ifnum\bozza=1
\newdimen\shift \shift=-2truecm
\def\lb#1{%
{\label{#1}\rlap{\kern\shift{$\scriptstyle#1$}}}}
\else\def\lb#1{\label{#1}} \fi

\begin{document}
\title{Optical excitation of phase modes in strongly disordered
  superconductors}
\author{T. Cea}
\affiliation{ISC-CNR and Dep. of Physics, ``Sapienza'' University of
  Rome, P.le A. Moro 5, 00185, Rome, Italy}
\author{D. Bucheli}
\affiliation{ISC-CNR and Dep. of Physics, ``Sapienza'' University of
  Rome, P.le A. Moro 5, 00185 Rome, Italy}
\author{G. Seibold}
\affiliation{Institut F\"ur Physik, BTU Cottbus, PBox 101344, 03013 Cottbus-Senftenberg,
Germany}
\author{L. Benfatto}
\affiliation{ISC-CNR and Dep. of Physics, ``Sapienza'' University of
  Rome, P.le A. Moro 5, 00185, Rome, Italy}
\author{J. Lorenzana}
\affiliation{ISC-CNR and Dep. of Physics, ``Sapienza'' University of
  Rome, P.le A. Moro 5, 00185, Rome, Italy}
\author{C. Castellani}
\affiliation{ISC-CNR and Dep. of Physics, ``Sapienza'' University of
  Rome, P.le A. Moro 5, 00185, Rome, Italy}
\date{\today}

\begin{abstract}
According to the Goldstone theorem the breaking of a continuous U(1)
symmetry comes along with the existence of low-energy collective modes. 
In the context
of superconductivity these excitations are related to the phase
of the superconducting (SC) order parameter and for clean systems are 
optically inactive. Here we show that for strongly disordered superconductors
phase modes acquire a dipole moment and appear as a subgap
spectral feature in the optical conductivity. This finding is 
obtained with both a gauge-invariant random-phase approximation scheme
based on a fermionic Bogoliubov-de Gennes state as well as with a prototypical
bosonic model for disordered superconductors. In the strongly disordered
regime, where the system displays an effective granularity of the SC
properties, the optically
active dipoles are linked to the isolated SC  islands,  offering a new
perspective for realizing microwave optical devices.

\end{abstract}


\maketitle

\section{Introduction}

In the last decades the failure of the BCS paradigm of
superconductivity in several materials led to a profound
modification of the description of the superconducting (SC) phenomenon
itself.  A case in point is the occurrence of Cooper pairing and phase
coherence at distinct temperatures, associated respectively with the
appearance of a single-particle gap $\D$ and a non-zero superfluid
stiffness $D_s$.  This behavior is observed, e.g., in high-temperature cuprate
superconductors \cite{review_lee,gomes07}, strongly-disordered films of
conventional
superconductors \cite{sacepe09,mondal11,sacepe11,chand12,noat13,pratap13}
and recently also in SC heterostructures \cite{mannhart_nat13}. In all
these materials the BCS prediction that
$D_s$ is of order of the Fermi energy, much larger than $\Delta\sim
T_c$, is violated due to the strong suppression of $D_s$.  The resulting
scenario, supported by systematic
tunneling measurements, suggests that pairing survives above $T_c$,
leading to a pseudogap state dominated by phase fluctuations 
enhanced by the low $D_s$ value.~\cite{note1}

In all this, optics represents a preferential playground to address the
peculiar role of disorder. 
Indeed, as we show in this Communication, disorder renders collective modes -
optically inactive in a clean superconductor - visible.  By analyzing
a prototype {\em fermionic} model, the attractive Hubbard model with on-site
disorder\cite{randeria01,dubi07,dubi08,erez10,randeria11,seibold12}, we
reveal that thanks to the breaking of translational invariance the collective
modes couple to light via an intermediate particle-hole excitation
process.  Most remarkably, this coupling leads to the emergence of
additional optical absorption, mainly due to phase modes, \emph{below}
the BCS-like threshold for a photon to break apart a Cooper pair, in
agreement with recent experimental
observations\cite{armitage_prb07,frydman_cm13}. 

Deeper insight into the nature of this disorder-induced optical response
is then gained through a comparison with the XY model in transverse random field.
Within this effective \emph{bosonic} description of disordered
superconductors\cite{ma_prb85,micnas_prb81}  we show explicitly how 
the local inhomogeneity of the
superfluid stiffness leads to a finite electric
dipole for the phase modes.
At strong disorder, where the
system segregates into SC islands of tens of
nanometers,\cite{randeria11,seibold12,sacepe11,pratap13} 
the SC dc current flows along preferential 
percolative paths through the good SC regions\cite{seibold12}. As a consequence the
finite-frequency optical absorption occurs in the remaining isolated SC regions,
thanks to the presence of a finite phase difference between the opposite sides of the
islands, which then act as  nano-antennas. This nano-scale selective optical effect, that we propose to test via microscopic
imaging\cite{shen_nano09}, can be used to tune the resonant frequency
and the quality factor of superconducting microresonators.\cite{zmuidzinas12}

\begin{figure}[h.tb]
\includegraphics[width=8.5cm,clip=true]{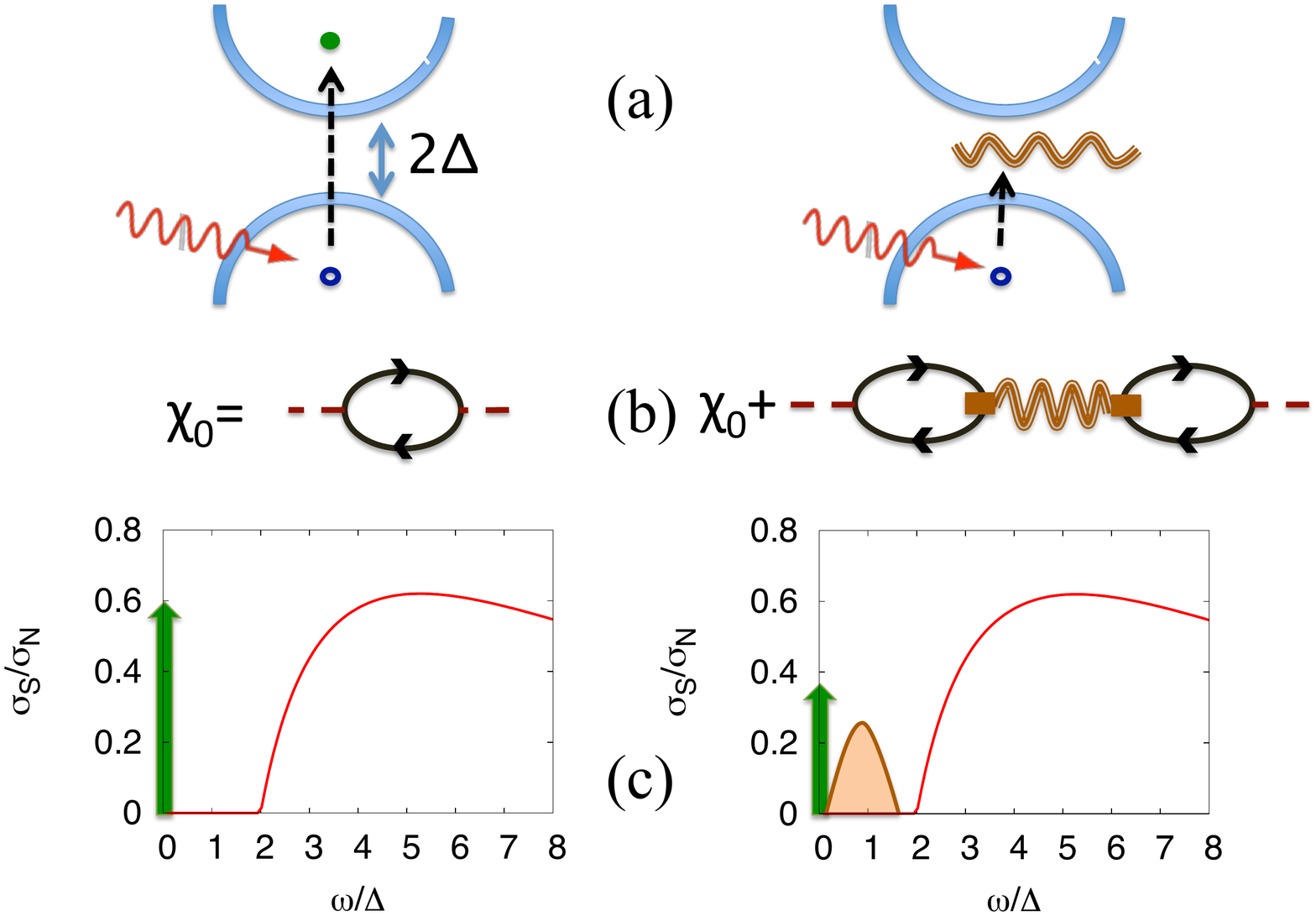}
\caption{Schematic of the optical absorption $\s_S/\s_N$ (S=SC,
  N=normal state) in a disordered
  superconductor. In the BCS approach (left) only the
  single-particle excitations across the SC gap $2\Delta$ (a) are
  included, corresponding to the bare-bubble approximation (b) for the
  current-current response function. The resulting optical
  conductivity (c) consists of a delta peak at $\o=0$ of weight $D_s^{BCS}$ (arrow) 
  plus a regular part (solid line) starting at $\o=2\D$. When vertex
  corrections are included (right) an excited quasiparticle can be
  converted in a
  collective mode (a), described in the diagrammatic
  approach (b) by the RPA resummations of the corresponding amplitude,
  phase or density fluctuations. An additional absorption appears at 
  energies $\o<2\D$ (c), corresponding to a superfluid peak at $\o=0$
  with strength $D_s<D_s^{BCS}$.}
\label{fig-diagrams}
\end{figure}

\section{Fermionic model}

The model Hamiltonian we consider to investigate a disordered
 superconductor is the attractive Hubbard model ($U<0$) with local
 disorder  $V_i \in  [-V,V]$ and hopping $t$ restricted to nearest-neighbors,
\begin{equation} 
\lb{ham}                                        
H=-t\sum_{\langle ij\rangle \sigma}(c^\dagger_{i\sigma}c_{j\sigma}+h.c.) + U\sum_{i}n_{i\uparrow}n_
{i\downarrow} +\sum_{i\sigma}V_i n_{i\sigma}                                      
\end{equation}                                                                    
which we solve in mean-field on a $N\equiv N_x\times N_y$ lattice
(up to $N=20\times 20$ with periodic boundary conditions) 
by using the BdG approach\cite{schrieffer,degennes}.
The total current 
in direction $\alpha$ is defined as usual as:
\bea
\lb{defcurr}
 J_\a(\bq,\o)&=&-e^2K_{\a\b}(\bq,\o)A_\b(\bq,\o),\\
\lb{defk}
K_{\a\b}(\bq,\o)&=&D\delta_{\a\b}-\chi_{\a\b}(\bq,\o)\, .
\eea
Here $D=\frac{-t}{N}\sum_{n,\s}\Big\langle\big(c_{n,\sigma}^\dagger c_{n+\hat\a\sigma} +h.c.\big)\Big\rangle
$ is the diamagnetic term, where $\langle \dots \rangle$ denotes the
thermal and disorder average, which restores the translational
invariance  for model \pref{ham}, allowing one to define the
Fourier transform $\chi_{\a\b}(\bq,\o)$
of the correlation function for the paramagnetic current $j_n^\alpha =
-i t \sum_\sigma\left\lbrack c_{n\sigma}^\dagger c_{n+\hat\alpha,\sigma} 
-c^\dagger_{n+\hat\alpha,\sigma}c_{n\sigma}\right\rbrack $. In a superconductor 
the superfluid stiffness is defined by the transverse $\bq \ra 0$
limit of Eq.\ \pref{defcurr}. For example, for a field along the $x$
direction one has $J_x=-e^2D_s A_x$ where
\bea
\lb{defds}
D_s&=&D-\mathrm {Re}\, \chi_{xx}(q_x=0,q_y\ra 0,\o=0).
\eea
The optical
conductivity is obtained from Eq.\ \pref{defcurr} by
assuming a homogeneous vector potential, so that $A_x(\o)=E_x(\o)/i\o$
and the real part of the optical conductivity is $\sigma(\o)=-e^2 \mathrm {Re}\,
\frac{K_{xx}(\bq=0,\o)}{i(\o+i0^+)}$, leading to
%
\bea
\s(\o)&=&e^2\pi\d(\o)\left[D-\mathrm
  {Re}\,\chi_{xx}({\bf 0},\o)\right]+e^2
\frac{\mathrm {Im}\, \chi_{xx}({\bf 0},\o)}{\o}\nn\\
\lb{defsigma}
&\equiv& e^2\pi D_s \d(\o)+\sigma_{reg}(\o)
\eea
where we separeted explicitly the superfluid response at $\o=0$ from
the regular part $\sigma_{reg}$ occurring at finite frequency. By  
using the Kramers-Kronig relations for the $\chi_{xx}$ one then finds
the well-know optical sum rule
\be
\lb{sumrule}
\int_0^\infty d\o \s(\o)=
\frac{\pi e^2}{2} D_s +\int_{0^+}^\infty \sigma_{reg}(\o)=
\frac{\pi e^2}{2} D.
\ee
The above Eq. \pref{sumrule} shows that any paramagnetic process
described by $\s_{reg}$ leads to a suppression of
the superfluid stiffness with respect to the diamagnetic term $D$,
which at small density and weak interactions reduces to the usual
form $D\simeq n/m$.
In the BCS theory $\chi\equiv\chi_0$ is computed in the
so-called bare-bubble approximation (see Fig.\ \ref{fig-diagrams}b,
left)\cite{schrieffer}, in which one includes only particle-hole excitations on
top of the BCS ground state. At $T=0$ these excitations are exponentially
suppressed by the opening of the gap, so that the optical absorption is
possible only above the threshold to break a
Cooper pair, i.e. at $\o>\o_{pair}=2\Delta$ (see Fig.\ \ref{fig-diagrams}c,
left). Provided that
$\o_{pair}$ is smaller than the inverse lifetime of
quasiparticles the resulting $\sigma^{BCS}_{reg}(\o)$ is given
by the well-known Mattis-Bardeen formula\cite{mattis}, and the superfluid
stiffness $D_s^{BCS}$ is smaller than $D$ already at $T=0$. In the
following we will show that also collective modes, neglected in the
BCS approach, give rise to a finite contribution to
$\sigma_{reg}(\o)$ at strong disorder, that is located mainly below
$\o_{pair}$ (see Fig.\ \ref{fig-diagrams}c,
right). This additional
optical absorption is accompanied by a further reduction of $D_s$ with
respect to $D_s^{BCS}$ \cite{seibold12}, that has been experimentally
reported\cite{mondal11,driessen_prl12,driessen_prb13}.

The full optical response beyond BCS level can be computed by
including vertex corrections\cite{schrieffer}, that also guarantee
full gauge invariance of the theory\cite{schrieffer,nota_ea}.
The current-current correlation function $\chi$ can then be
expressed in a compact form as (see App. A):
\be
\lb{chirpa}
\chi=\chi_0+\hat \L^T V [1-\hat \Pi^0 V]^{-1}\hat \L
\ee
where $\hat \L$ is the vector containing the correlation functions
that couple the current $j^\a_n$ to collective modes, i.e. 
particle-particle (amplitude and phase) and density
fluctuations, described by the 
RPA resummation of the bare susceptibility  $\hat \Pi^0$, see Fig.\
\ref{fig-diagrams}b right. The $\hat V$ and $\hat\Pi^0$ are matrices both in real space and in the
phase space of collective modes, and translational invariance for
$\chi$ is
recovered after average over disorder configurations.
In the clean case collective modes contribute only to the longitudinal
response at
finite $\bq$ \cite{schrieffer,nota_ea}. In contrast,
disorder renders the $\hat\L$ susceptibilities finite even for a
$\bq=0$ external perturbation, so that
the collective modes contribute to the optical response. 
Notice that this optical mechanism is similar to the one
discussed recently for few-layer graphene\cite{fano-rice}
to explain the huge infrared-phonons peaks\cite{fano-exp}.
In that case  doping activates the intermediate particle-hole process,
analogously to what disorder does in our problem.

\begin{figure}[htb]
\includegraphics[width=8.5cm,clip=true]{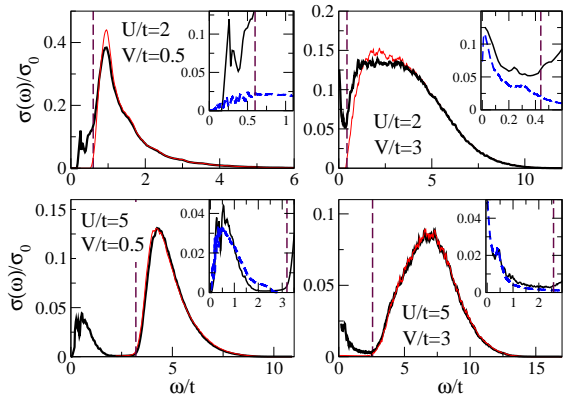}
\caption{ 
  $\sigma(\omega)$ in units of $\s_0\equiv e^2/\hbar$ 
  for the fermionic model \pref{ham}. Here $N=$ and we averaged over
  ... disorder configurations. The main panels report
the curves without (thin, red) and with (thick, black) vertex
corrections, while the dashed vertical lines mark
$\o_{pair}=2\Delta$. Insets: zoom of the low-energy part along with
the results of the bosonic model \pref{hamps} (dashed lines), with $W,J$ parameters
assigned as described in the text.}
\label{fig-sigma}
\end{figure}

\begin{figure}[htb]
\includegraphics[width=8cm,clip=true]{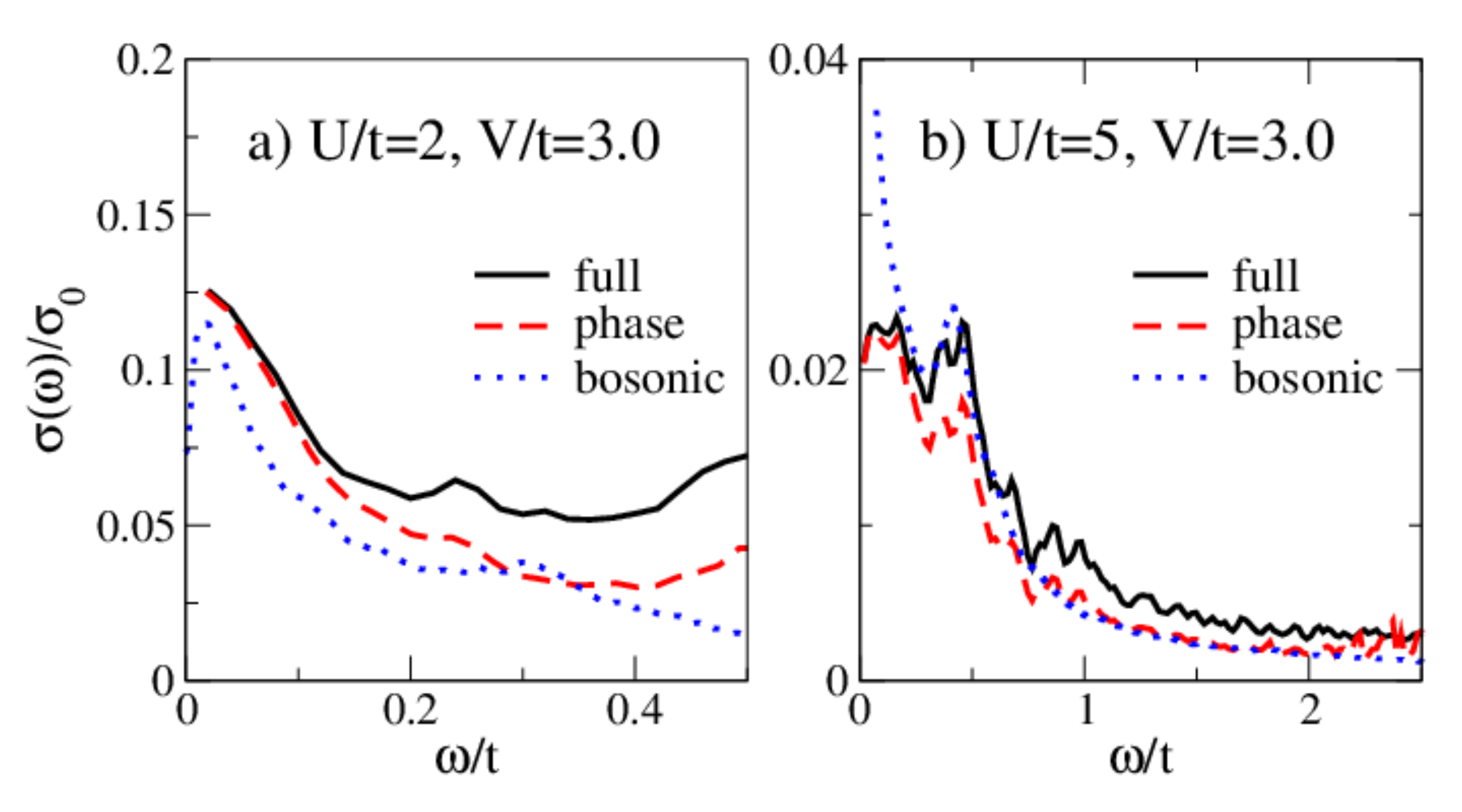}
\caption{Comparison between $\s(\o)$ computed using the full
  gauge-invariant response (solid line, black) and the contribution 
of phase fluctuations only (dashed line, red) (see Appendix A). 
The results of Fig.\ \ref{fig-sigma} for the bosonic model \pref{hamps}
(dotted line, blue) are reported as well.}
\label{fig-ponly}
\end{figure}

The results for the optical conductivity at finite frequency for two
representative values of coupling $U$ and disorder are shown in Fig.\
\ref{fig-sigma}, along with their BCS counterparts. As one can see,
the major differences between the two appear below the scale
$\omega_{pair}=2\Delta$, marked with a dashed line. Notice that in the
model \pref{ham} the spectral gap $\Delta$ in the single-particle
excitations remains finite (and relatively large) at strong disorder,
as it has been discussed
previously\cite{randeria01,dubi07,randeria11}. As a consequence, the
BCS calculation always shows a finite threshold at $\o_{pair}$, with a
profile that coincides at low disorder with the Mattis-Bardeen
prediction\cite{mattis}. In contrast, the full response extends also
below $\omega_{pair}$, with a shape and intensity that depend both on
the SC coupling $U$ and disorder. This result can explain the residual
optical absorption in the microwave
regime\cite{armitage_prb07,frydman_cm13} and deviations from BCS
theory\cite{mondal11,driessen_prl12,driessen_prb13} observed recently
at strong disorder. In
particular, the smearing of the $\o_{pair}$ treshold due the
 presence of a dissipative channel
associated to phase modes can
lead to an apparent optical gap smaller
than the one measured by STM, explaining the puzzling results of Ref.\
[\onlinecite{frydman_cm13}]. At the same time this effect can influence the performance of
superconducting microwave devices, a field that has grown dramatically
over the past decade\cite{zmuidzinas12}. Finally, two remarks are in
order with respect to the results of Fig. \ref{fig-sigma}. First of
all, we checked that even 
though all the collective modes enter in the full response, the main
contribution to $\s_{reg}(\o)$ at low energy stems from phase
fluctuations. This is shown explicitly in Fig.\ \ref{fig-ponly}, where
$\sigma_{reg}(\o)$ has been computed by including only the
phase-current vertex in Eq. \ref{chirpa} (see Appendix A).
Second, one could wonder what happens when the 
Coulomb interaction, neglected in the present calculations, is taken
into account. Indeed, one usually expects that the presence of 
long-range interactions the sound-like dispersion of 
phase modes is converted into a plasmonic one.\cite{nota_ea} 
However, we
do not expect that this result will spoil our conclusions: indeed,
while plasmonic modes appear in the longitudinal response,  the
transverse optical response, which is the one discussed here, should
be anyway screened, as shown explicitly in
the weakly-disordered case in Ref.\ [\onlinecite{belitz}].

\section{Bosonic model}
To systematically address the structure of the phase excitations
responsible for the sub-gap absorption we compute the optical
conductivity also within an effective bosonic model for the disordered
superconductor, i.e.  the $XY$ spin 1/2 model in a transverse random
field\cite{ma_prb85}:
\begin{equation}
\lb{hamps}
\mathcal{H}_{PS}\equiv -2\sum_i\xi_iS_i^z-2J\sum_{\langle i,j\rangle}\left(S^+_iS^-_j+h.c.\right).
\end{equation}
In the pseudo-spin language $S^{z} = \pm 1/2$ corresponds to a site occupied
or unoccupied by a Cooper pair, while superconductivity corresponds to
a spontaneous in-plane magnetization, e.g. $\langle
S^x_i\rangle\neq 0$. 
Disorder is represented by the random transverse field
$\xi_i$, box distributed between $-W$ and $W$. The optical response of
classical\cite{stroud_prb00} and quantum\cite{randeria13}  $XY$-like
models has been addressed previously, by introducing disorder in the coupling
$J$. The 
model \pref{hamps} focuses instead on the
competition between pair hopping ($J$) and
localization ($W$)\cite{ma_prb85,fim_prb10,lemarie_prb13,laflorencie_prl13}, 
that has been recently
proven successfull\cite{sacepe11,lemarie_prb13} to describe the STM experimental results in the 
SC phase near the SIT. 
We first solve the model (\ref{hamps}) in mean-field to determine 
$\langle
S^x_i\rangle=\frac{1}{2}\sin\theta_i$ 
and then rotate to the local coordinate system such that 
the new  z-axis is $\widetilde{S_i^z}= S^z_i\cos\theta_i+S^x_i\sin\theta_i$.
At strong disorder the SC order
parameter develops an inhomogeneous spatial
distribution, with SC islands
embedded in an insulating background (cfr Fig.\ \ref{fig-curr} below), in analogy both with the
fermionic model \pref{ham}\cite{randeria01,dubi07,seibold12} and with
tunnelling experiments\cite{sacepe11,pratap13}.  Small fluctuations
with respect to the mean-field configuration can be described by means of a
Holstein-Primakov (HP) scheme, where spins are bosonized as usual as
$\widetilde{S_i^z}=1/2-a^+_ia_i$, $\widetilde{S_i^+}\simeq a_i$ and 
$\widetilde{S_i^-}\simeq a^+_i$. Here we have defined
$\widetilde{S_i^\pm}=\widetilde{S_i^x}\pm i \widetilde{S_i^y}$ with 
$\widetilde{S_i^x}= -S^z_i\sin\theta_i+S^x_i\cos\theta_i$ and
$\widetilde{S_i^y}= S^y_i$ being orthogonal to the
local quantization axis. The Hamiltonian \pref{hamps} is then mapped
into a quadratic model $\mathcal{H}_{PS} =E_{MF}+\mathcal{H}'_{PS}$  that can be diagonalized by means of a
Bogoliubov transformation $a_i=\sum_\alpha \left(u_{\alpha
    i}\gamma_\alpha +v_{\alpha
    i}\gamma^\dagger_\alpha\right)$:
\bea
\mathcal{H}'_{PS}&=&\sum_{ij}\left[	A_{ij}(a^\dagger_ia_j+h.c.)
  +\frac{1}{2}B_{ij}\left(	a_ia_j+h.c.\right)
\right]\nn\\
\lb{H_harm}
&=&\sum_\alpha E_\alpha \gamma^\dagger_\alpha \gamma_\alpha+const.
\eea
Here $A_{ij}=2\delta_{ij}\xi_i/\cos \theta_i-J(1+\cos\theta_i\cos\theta_j)(1-\d_{ij})$ and
$B_{ij}=J(1-\cos\theta_i\cos \theta_j)(1-\delta_{ij})$ are the
matrices that enter in the eigenvalue problem for the
excitation energies $E_\a$\cite{nota_gm}. 
The equivalence between the HP excitations and the SC phase excitations at
Gaussian level
can be made explicit by the identification of the phase operators
$\Phi_i$ and their conjugated momenta $L_i$,
\bea
\lb{phiop}
\Phi_i&=& -
2\frac{S^y_i}{\sin\theta_i}=\sum_\alpha i \frac{\phi_{\a i}}{\sqrt 2}
(\g_\a^\dagger-\g_\a),\\
L_i&=&S_i^\perp \sin \theta_i=\sum_\a \frac{\ell_{\a i}}{\sqrt 2}
(\g_\a^\dagger+\g_\a),
\eea
where $\phi_{\a i}=\sqrt{2}\left(v_{\alpha
    i}-u_{\alpha i}\right)/\sin\theta_i$ and $\ell_{\a i}=\left(u_{\alpha
    i}+v_{\alpha  i}\right)\sin\theta_i/\sqrt{2}$. The fluctuation part of the
Hamiltonian \pref{H_harm} can then be expressed as:
\be
\lb{hpf}
{\cal H}'_{PS}=\frac{1}{2}\sum_{i,\mu= x, y} J^\mu_i\left[ \Delta_\mu
  \Phi_i\right]^2 +\frac{1}{2}\sum_{ij} {\cal X}^{-1}_{ij} L_i L_j
\ee
where $J^\mu_i\equiv  J\sin \theta_i \sin \theta_{i+\hat\mu}$ are the local
stiffnesses of the disordered superconductor, $\D_\mu$ is the
discrete derivative in the $\mu$ direction and ${\cal
  X}^{-1}_{ij}=2(A_{ij}+B_{ij})/\sin \theta_i \sin \theta_j$ is the
inverse matrix of the compressibilities.  Consistently with the identification
\pref{phiop} the usual Peierls coupling to the
gauge field in the 
pseudospin model \pref{hamps} corresponds to the replacement $S^+_{i}S^-_{i+\mu}
\rightarrow S^+_{i}S^-_{i+\mu} e^{-2ieA_\mu}$, with a factor of 2 accounting
for the double charge of each Cooper
pair. This leads in Eq.\ \pref{hpf} to the shift
$\Delta_\mu  \Phi_i \ra \Delta_\mu  \Phi_i-2eA_\mu$, i.e. 
the usual minimal-coupling scheme. The real part of the optical conductivity
for the bosonic model \pref{hamps} is then easily obtained as
$\s^B(\o)=e^2 \pi \d(\o) D^B_s +\sigma_{reg}^B(\o)$ with
\bea
\lb{dsb}
D_s^B&=&D^B-\frac{1}{N}\sum_\a Z_\a\\
\lb{sigmab}
\s_{reg}^B(\o)&=&\frac{e^2\pi}{2N} \sum_\a
Z_\a\left[\d(\o+E_\a)+\d(\o-E_\a)\right],
\eea
where $D^B=(1/N)\sum_i 4  J^\mu_i$ is the diamagnetic term of the
bosonic model \pref{hamps}, $\mu=x$ for instance and the effective dipole $Z_\a$ of each
excitation mode is
\be 
\lb{za}
Z_\a=\frac{1}{E_\a}\left[\sum_i 2J_i^\mu \Delta_\mu
  \phi_{\a i}\right]^2.
\ee
\begin{figure}[htb]
\includegraphics[width=8cm,clip=true]{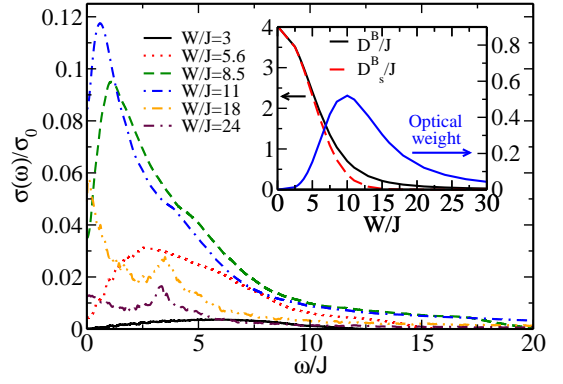}
\caption{ $\s_{reg}(\o)/\s_0$ for the bosonic model \pref{hamps} 
at different values of $W/J$. The lattice size is $N=50\times 50$ and the
average is taken over 100 disorder configurations. 
Inset: disorder
  dependence of the diamagnetic term
$D^B$, the superfluid stiffness $D_s^B$ and the total spectral
weight   $\int_{0^+}^\infty d\o\s_{reg}(\o)=(\pi/2) (D^B-D_s^B)$ in
units of $J$. }
\label{fig_opt-bos}
\end{figure}
For a uniform stiffness ($J_i^\mu=$const) one finds that $Z_\a$ is 
proportional to the total derivative of the phase
modulation, which then vanishes for periodic boundary conditions. Thus, the
inhomogeneity of $J^\mu_{i}$ induced by disorder is a crucial
prerequisite to obtain a finite electric dipole, responsible for the 
$\s_{reg}(\o)$ shown in Fig.\
\ref{fig_opt-bos}. As one can see, the 
optical response moves towards decreasing energies for increasing
disorder (i.e.  $W/J$), and its total spectral weight $\int_{0^+}^\infty
d\o \s_{reg}(\o)=(\pi/2)(D^B-D^B_s)$ [see Eq.\
\pref{dsb}-\pref{sigmab}] first increases, due to the disorder-tuned
optical absorption at finite $\o$, and then decreases again, due to
the strong suppression by disorder of the diamagnetic term $D^B$ itself 
(see inset Fig.\ \ref{fig_opt-bos}). Notice that the decrease
of $D^B$ with increasing disorder reflects the suppression of the
local order parameter, encoded in the
fermionic language \pref{ham} in the suppression of the BCS
stiffness $D_s^{BCS}$. This analogy can be used to obtain a
quantitative comparison between the fermionic and the bosonic
approach, by fixing $W/J$ of the model \pref{hamps} in order to reproduce 
$D^B_s/D^B=D_s/D_s^{BCS}$. In this way we can account in both 
models for the same transfer of spectral weight from $\o=0$ to
$\s_{reg}(\o)$ (see Appendix B). 
The results are shown in the insets of
Fig.\ \ref{fig-sigma} and in Fig.\ \ref{fig-ponly}: as one can see, 
at large $U$ the
bosonic model reproduces in a quantitative way the characteristic
energy scales for optical absorption in the fermionic model. At weaker 
coupling the comparison is instead only qualitative, due partly to the
difficulties of clearly separating the contribution of quasiparticles
and collective modes. 

\begin{figure}[htb]
\includegraphics[width=8cm,clip=]{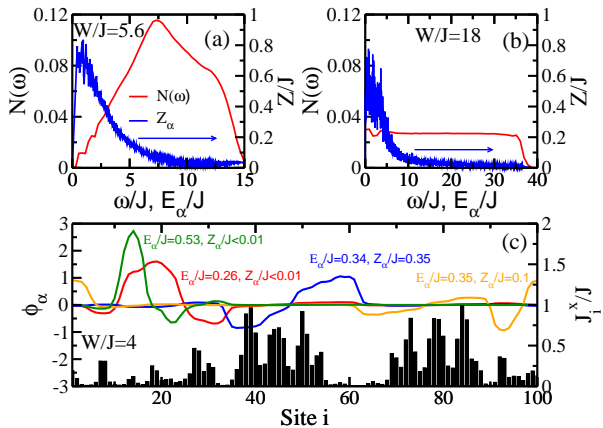}
\caption{ (a)-(b) Density of phase modes $N(\o)$ and effective dipole
  $Z_\a$ at two values of disorder (averaged over 100 disorder configurations). (c) Spatial
  structure of the phase modes in the one-dimensional case at selected
  energies for a given disorder realization, along with the spatial variations of the stiffness $J_i^x$,
  represented with bars. The  largest effective dipole is realized for the blue and
  orange excitations, whose monotonic phase variations overlap with a 
  region of large local stiffness.}
\label{z-dos}
\end{figure}


Let us finally analyze the connection between the optical response and the
inhomogeneous spatial distribution of the SC properties.
The optical response \pref{sigmab} 
is proportional to the density of states of phase modes $N(\o)$,
weighted by the effective dipole function $Z_\a$ of Eq.\
\pref{za}. Both quantities depend on disorder, as it is shown in Fig.\
\ref{z-dos}a,b, and in general the $1/E_\a$ prefactor of Eq.\
\pref{za} favors a larger dipole for lower-energy modes. In addition at
strong disorder, when the system segregates into SC islands with large
local stiffness $J_\mu^i$, the optical absorption is large
when the phase excitations occur
{\em inside} the SC regions, according to Eq.\ \pref{za}. 
This effect can be better visualized
in a one-dimensional version of the model \pref{hamps}, as it is shown
in panel (c) of Fig.\ \ref{z-dos}. Here one can clearly see that the
largest optical dipole is realized when a monotonic phase variation
overlaps with a good SC region. Since the charge is the conjugate
variable of the phase gradients, one then realizes a charge unbalance
on the two sides of the island, making it optically active. 

\begin{figure}[htb]
\includegraphics[width=8.5cm,clip=]{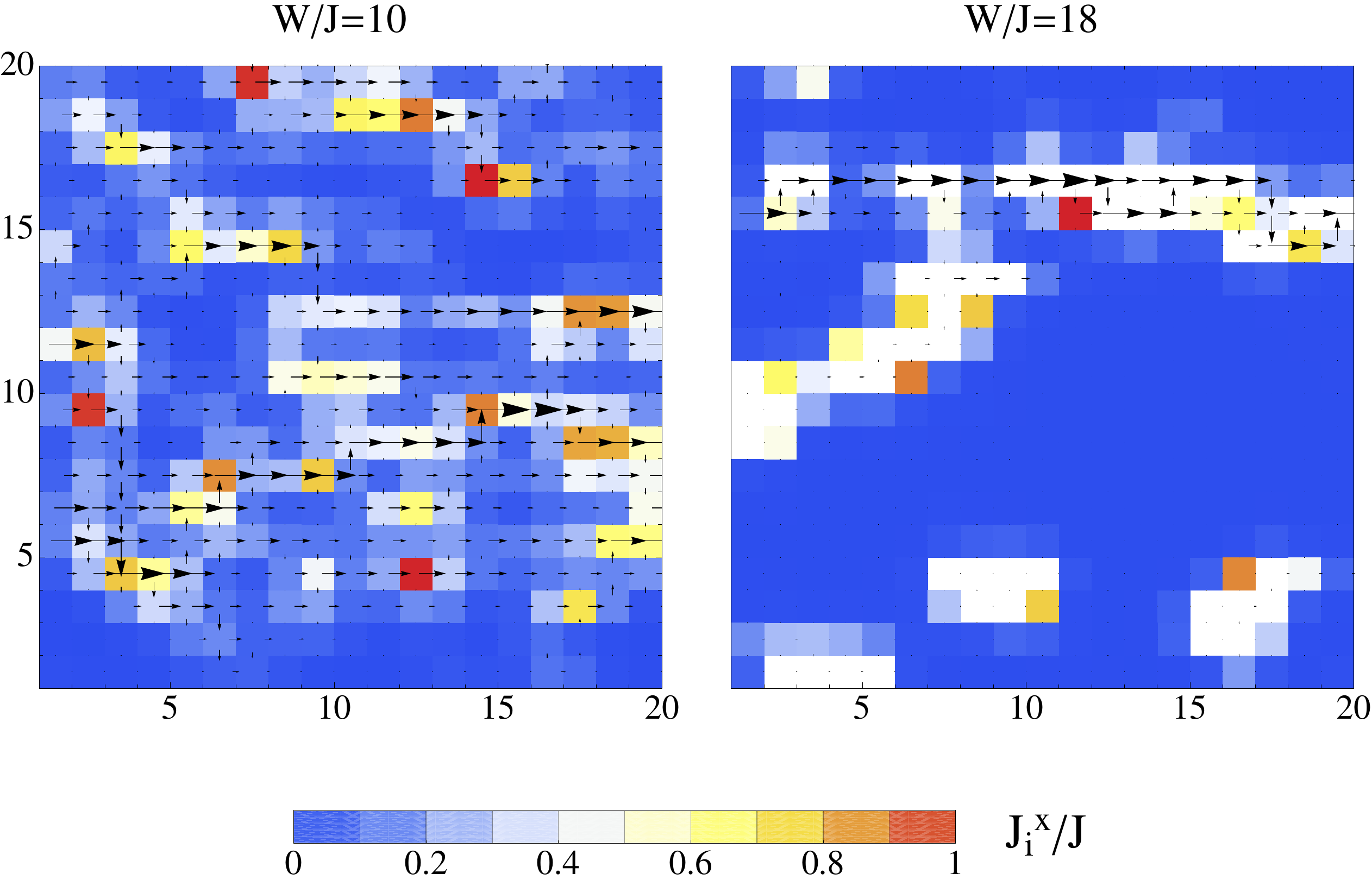}
\caption{ Local supercurrents (arrows) for an applied field ${\bf
    A}=-A \hat x$ superimposed over the map of the local stiffnesses
  $J_i^x/J$ for a given disorder realization and two values of
  $W/J$. The size of the arrows is proportional to the strenght of the total
  local current, whose diamagnetic contribution is proportional to the
  local stiffness displayed in the underlying map. The current flows along preferential paths connecting the
  regions with large $J_i^x$. This means that the isolated SC islands, i.e. those
  which reside far from the main percolative paths of the current,
  have a large paramagnetic response responsible for the 
absorption at finite frequencies. For
example, for $W/J=18$ all the diamagnetic contribution of the white
regions on the bottom of the map is transferred to $\s_{reg}(\o)$.} 
\label{fig-curr}
\end{figure}

At strong disorder this space-selective optical absorption is strictly connected to the
emergence of percolative paths for the superfluid
currents, analogous to the ones discussed in Ref.\
[\onlinecite{seibold12}] for the fermionic model \pref{ham}. In Fig.\
\ref{fig-curr} we show at two values of $W/J$ the currents  in the presence of a finite applied field ${\bf
  A}=-A\hat x$ for a given disorder configuration, superimposed to the
map of the local stiffnessess $J_i^x$. Since $J_i^x$ is a
measure of the local diamagnetic response, a small current
occurring over a good SC region is due to a large local
paramagnetic response, i.e. to an optical absorption at finite
frequencies. At strong disorder the percolative
supercurrent paths
leave aside several isolated SC islands, which then contribute to
$\s_{reg}(\o)$ thanks to the dipole-activation mechanism
explained above. 

\section{Conclusions}

In summary, we computed the optical response due to collective modes
in two prototype fermionic and bosonic models for disordered
superconductors. In both cases we find that disorder renders phase
fluctuations optically active, in a range of energies that lies below
the threshold for single-particle excitations for the fermionic
case. The bosonic approach allows us to establish 
a clear
correspondence between the optical response and the spatial
inhomogeneity of the SC order parameter, showing that 
optical absorption stems predominantly from phase fluctuations within
the good SC regions.  Besides explaining recent experiments in
strongly-disordered superconductors\cite{armitage_prb07,frydman_cm13,driessen_prl12,driessen_prb13}
our results could be further checked experimentally by means of
near-field scanning microwave impedance
microscopy\cite{shen_nano09}. Indeed, the proposed mechanism of
direct correspondence between the SC granularity and optical
absorption, evidenced in Fig.\ \ref{fig-curr}, can be potentially
mapped out by this technique, able to resolve spatial variations at
length scales well below the radiation wavelength. In this respect the
variation of the microwave optical properties of disordered superconductors at
the nanoscale can be used to improve the performance of
SC microresonators built in standard geometries, or even to design new
nano-electric devices targeted for space- and frequency-selective
applications.

We thank E. Driessen, A. Frydman and D. Sherman for useful discussions.
L.B. acknowledges financial support by MIUR under Grant
No. FIRB2012 (RBFR1236VV). 

\appendix

\section{Optical conductivity for the fermionic model}

In order to compute fluctuations on top of the inhomogeneous BdG ground state 
we evaluate dynamical  correlation functions
\begin{equation}
\chi_{nm}(\hat{X},\hat{Y})=-i\int\!dt e^{i\omega t}\langle {\cal T} \hat{X}_n(t) \hat{Y}_m(0)\rangle
\end{equation}
where $\hat{X}$,$\hat{Y}$ correspond
to either pair or charge fluctuation operators,
i.e. $\delta\Delta_i \equiv c_{i\downarrow}c_{i\uparrow} - \langle
c_{i\downarrow}c_{i\uparrow} \rangle_0 $, $\delta\Delta_i^\dagger
\equiv c^\dagger_{i\uparrow}c^\dagger_{i\downarrow} - \langle
c^\dagger_{i\uparrow}c^\dagger_{i\downarrow} \rangle_0$ and 
$\delta n_i \equiv
 \sum_{\sigma}\left(c^\dagger_{i\sigma}c_{i\sigma} - \langle
   c^\dagger_{i\sigma}c_{i\sigma} \rangle_0\right) $. 
Here and in the following a nought sub- or superscript denotes evaluation in 
the BdG ground state.

The local interactions between
pair ($U\delta\Delta_i^\dagger\delta\Delta_i$, 
$U\delta\Delta_i\delta\Delta_i^\dagger$) and charge 
($U/2\delta n_i\delta n_i$) fluctuations are 
contained in the matrix $\underline{\underline{V}}$ so that
 the resummation 
\begin{equation}
\underline{\underline{\chi}}=\left\lbrack \underline{\underline{1}} -
\underline{\underline{\chi^0}}\underline{\underline{V}}\right\rbrack^{-1}
\underline{\underline{\chi^0}}
\end{equation}
allows to compute the dynamical 
amplitude $A_i\equiv (\delta\Delta_i+\delta\Delta^\dagger_i)/\sqrt{2}$ and
phase $\Phi_i\equiv (\delta\Delta_i-\delta\Delta^\dagger_i)/\sqrt{2}$
correlations.

Vertex corrections to the bare current-current 
correlation function $\chi^0_{nm}(j^\alpha,j^\beta)$ can be obtained
by defining $\Lambda_{nm}^\alpha=\chi^0_{nm}(j^{\alpha},\hat{Y})$
and $\overline{\Lambda}_{mn}^\alpha=\chi^0_{mn}(\hat{Y},j^{\alpha})$
which couple the current $j^{\alpha}_n$ between sites 
$R_{n}$ and $R_{n+\hat\alpha}$ to the pair and charge fluctuations
$\hat{Y}_m\equiv(A_m,\Phi_m,\delta n_m)$.
The full (gauge invariant) current correlation function is then obtained from
\begin{eqnarray}
\chi_{nm}(j^\alpha_n,j^\beta_m)&=&\chi^0_{nm}(j^\alpha_n,j^\beta_m)
+ \Lambda_{nm}^\alpha V_{mk}\overline{\Lambda}_{km}^\beta \nonumber \\
&+& \Lambda_{nm}^\alpha V_{mk} \chi_{kl}V_{ls}\overline{\Lambda}_{sm}^\beta \nonumber \\
&=& \chi^0_{nm}(j^\alpha_n,j^\beta_m)\nonumber \\
 &+& \Lambda_{nm}^\alpha V_{mk}
\left\lbrack \underline{\underline{1}} - \underline{\underline{\chi^0}}\underline{\underline{V}}\right\rbrack^{-1}_{kl}\overline{\Lambda}_{lm}^\beta \,. \label{eqs3}
\end{eqnarray}
The average over several disorder configurations (typically $50 -70$) 
restores translational
invariance for $\chi_{nm}$, so that $\sigma(\o)$ can be computed
according to Eq.\ \pref{defsigma}. The predominant role of phase
fluctuations for the subgap optical response is demonstrated in 
Fig. \ref{fig-ponly}. Here the dashed lines correspond to $\s(\o)$
computed by including only the phase-current vertex in Eq. \ref{eqs3},
whereas the solid lines correspond to the full optical conductivity,
dressed by all the collective modes.

\section{Comparison between the bosonic and fermionic model}

In order to make a  quantitative comparison between the bosonic and
fermionic models we propose a scheme based on the equivalence of the 
optical spectral weight due to collective modes in 
both models. In the
fermionic model the diamagnetic term $D$ is only weakly dependent on
disorder. On the other hand,  the BCS estimate of the
superfluid density $D_s^{BCS}$ is strongly suppressed due to the enhanced 
paramagnetic contribution of quasiparticles.
In the bosonic model quasiparticle excitations are not
present: however, the localization effects due to disorder are taken
into account in the effective local stiffnesses $J_i^\mu$, which enter
in the bosonic diamagnetic term $D^B$. Any additional correction due
to phase fluctuations is encoded in the ratio
$D_s/D_s^{BCS}$ or $D^B_s/D^B$ for the fermionic or bosonic model,
respectively.  
The ratio $D_s/D_s^{BCS}$ measures also the
relative strength of the sub-gap optical response with respect to the BCS-like
part. However, since this one is also slightly modified by vertex
corrections, especially at weak coupling (see red and black lines in
Fig.\ \ref{fig-sigma}), we estimate an effective $\tilde D_s^{BCS}$
from the integrated spectral
weight at $\o>2\D$ of the full optical conductivity. 
 Thus, for a given $U$ in the
fermionic model \pref{ham} we determine $W/J$ in the bosonic
model \pref{hamps} in order to have $D^B_s/D^B=D_s/\tilde D_s^{BCS}$. This
establishes a rescaling function $\alpha$ such that
$W/J=\alpha(V_0/t)$. 
We then fix a factor $\gamma$ ($J=\gamma t$) such that $D^B/J=\tilde D_s^{BCS}/\gamma
t$, and we then plot $\s_{reg}$ as a function of $\o/t$ as shown in
the insets of Fig.\ \ref{fig-sigma}. This procedure reveals that already for $U/t=5$,
i.e in the intermediate-coupling regime, the ratio $J/t$
obtained in this way is very similar to the value $J/t\sim t/U$ 
predicted by the exact mapping of the clean attractive Hubbard model onto the
pseudospin model\cite{micnas_prb81}.


\begin{thebibliography}{cc}
\bibitem{note1} Actually the origin of 
the pseudogap in underdoped cuprate superconductors is still an open
problem, since it may also 
involve a second energy scale originating from a 
competing order (cf. Ref.~\cite{gomes07}).
\bibitem{nota_gm}
We verified that the first eigenvalue
always corresponds to a zero-energy mode, as expected for the $\bq=0$
limit of the Goldstone mode of the $U(1)$ symmetry  broken in the SC phase.
\end{thebibliography}
\end{document}